\documentclass[prd,twocolumn,superscriptaddress,showpacs,nofootinbib,
preprintnumbers]{revtex4}

\usepackage{amsmath}
\usepackage{amsfonts}
\usepackage{graphicx}
\usepackage{dcolumn}

\def\be{\begin{equation}}
\def\ee{\end{equation}}
\def\ba{\begin{eqnarray}}
\def\ea{\end{eqnarray}}
\def\bs{\begin{subequations}}
\def\es{\end{subequations}}

\usepackage{color}


\begin{document}

\title{Reheating the D-brane universe via instant preheating }
\author{Sudhakar Panda}
\affiliation{Harish-Chandra Research Institute, Chhatnag Road,
Jhusi, Allahabad-211019, India}
\author{M. Sami}
\affiliation{Centre of Theoretical Physics, Jamia Millia Islamia,
New Delhi-110025, India}
\author{I. Thongkool}
\affiliation{Centre of Theoretical Physics, Jamia Millia Islamia,
New Delhi-110025, India}
\begin{abstract}
We investigate a possibility of reheating in a scenario of D-brane
inflation in a warped deformed conifold background which includes
perturbative corrections to throat geometry sourced by chiral
operator of dimension $3/2$ in the CFT. The effective D-brane
potential, in this case, belongs to the class of non-oscillatory
models of inflation for which the conventional reheating mechanism
does not work. We find that gravitational particle production is
inefficient and leads to reheating temperature of the order of
${10^8} GeV$. We  show that instant preheating is quite suitable to
the present scenario and can easily reheat universe  to a
temperature which is higher by about three orders of magnitudes than
its counter part associated with gravitational particle production.
The reheating temperature is shown to be insensitive to a particular
choice of inflationary parameters suitable to observations.

\end{abstract}

\maketitle
\section{Introduction}

Inflationary scenario in cosmology is known to solve a number of
problems associated with the big bang model \cite{inflation,linde}
and is well supported by observation\cite{Spergel1,Spergel2}.  It is
believed that universe, at the end of the inflationary epoch,
consisted almost entirely of the homogeneous inflaton field which
decayed into inhomogeneous fluctuations and other particles. This
decay period, known as reheating \cite{kls97}, is fairly understood
and has provided crucial links between the inflationary epoch and
the subsequent thermalized era. In the early days of the reheating
proposal, it was believed that the inflaton field decayed
perturbatively as a collection of particles and during the decay it went through a large number
of oscillations around the minimum of its potential. A different hypothesis is that the decay could have been initiated  by coherent
field effects such as parametric resonance. For such a decay process the field  undergoes only a  few
oscillations. This rapid non-perturbative decay has been termed as
preheating. Subsequently a new proposal for reheating, dubbed {\it
instant preheating} was made in Refs.\cite{FKL,FKL1} which does not
require oscillation of the inflaton field.

In this paper we investigate the instant preheating mechanism in a
recently studied brane inflation model \cite{acps} using the leading
corrections to the brane potential obtained in \cite{BDKKM}. The
model used a single throat, containing the  D3-brane moving towards
the ${\bar D}$3-brane located at the tip of the throat. The position
of the D3-brane with respect to the localized ${\bar D}$3-brane  is
interpreted  as an  inflaton field following the method outlined in
the original proposal of Ref.\cite{KKLMMT}. Earlier attempts to
study reheating in brane inflation model using multi-throat
configurations were made in Refs. \cite{FKL2,bbc,sm}.

We recall that the proposal in Ref. \cite{KKLMMT} is based on type IIB
string compactification, with background fluxes that stabilize all
the complex structure moduli fields \cite{GKP}, using a
Klebanov-Strassler (KS) throat \cite{KS}. It also made use of the
proposal in Ref.\cite{KKLT} for the stabilization of
$K{\ddot{a}}$hler moduli fields due to nonperturbative effects that
arise via the the gauge dynamics of either an Euclidean D3-brane or
a stack of D7-branes wrapping a four cycle in the warped throat. The
AdS vacuum is lifted to a de-Sitter vacuum by placing ${\bar
D}$3-brane at the tip of the throat since the warped tension of the
anti-brane adds to the energy density of the system. Inflation, in
this scenario, is realized by the motion of a D3-brane towards the
anti-brane due to their mutual Coulomb attraction and the radial
separation between them serves as the inflaton field. However, the
moduli stabilization, in presence of the mobile D3-brane affects the
potential for the inflaton field  yielding a mass term for the
inflaton which, unfortunately, turns out to be of the order of the
Hubble parameter and hence effectively spoils the inflation. To
bring out new features of the inflaton potential, it was proposed
\cite{Bau2,Bau3,KP} to restrict the motion of D3-brane deep inside
the throat  assuming certain embeddings of the stack of D7-branes. An inflationary dynamics was studied in \cite{PSS} using the potential obtained in the above model, which involves two scalar fields, namely, the volume modulus field and the  field representing the radial distance between the brane and the anti-brane. The result turned out to be unrealistic since not only
it needed severe fine tunings but also it was found that in case of
 the scale invariant spectrum of scalar perturbations, their amplitude becomes much larger than
the COBE normalized value (see, Ref.\cite{HC} on the related theme).

A  proposal, using the holographic point of view, to account for the ultraviolet physics arising
from gluing the warped throat to the compact Calabi-Yau manifold is
made in Ref.\cite{BDKKM}. The authors employed gauge/string
correspondence for the warped deformed conifold. In this framework,
the position of the mobile D3 brane is identified with the Coulomb
branch vacuum expectation value of the dual gauge theory. In such a
holographic formulation, the bulk effects are encoded in the
coupling of a gauge invariant operator ${\cal O}_{\Delta}$ of
scaling dimension $\Delta$ to its dual bulk mode in the gauge theory
which modifies the inflaton potential with an additional term given by

\begin{equation}
V_{\Delta}~=~- c a_0^4 T_3 \left(
\frac{\phi}{\phi_{UV}}\right)^{\Delta},
\end{equation}

where $c$ is a positive constant and $a_0$ is the warp factor at the
tip of the deformed conifold which depends upon the background
fluxes. The potential above is obtained after the minimization over
the angular location and written in terms of the canonically
normalized inflaton field $\phi = \sqrt{T_3} r$, where $ r$ is the
radial distance of the D3-brane from the tip of the throat, $T_3$
being the tension of the D3-brane  and $\phi_{UV}$ is the value of
the the field $\phi$ corresponding to a cut off scale $M_{UV}$ of
the conformal field theory. Moreover, it was  found that the leading
corrections to the inflaton potential comes either from modes
corresponding to $\Delta = 3/2$ related to the scaling dimension of
a chiral operator in the gauge theory or $\Delta = 2$ corresponding
to the same of a non-chiral operator. If the former is not forbidden
by a certain discrete symmetry  of the string compactification then
this operator provides the dominant correction to the potential. We
consider this situation and include the other known contribution to
the inflaton potential after which the full potential is found to be

\begin{equation}
V (\phi )~=~ D \left[ 1 + \frac{1}{3} \left(
\frac{\phi}{M_{pl}}\right)^2 - C_{3/2} \left(
\frac{\phi}{M_{UV}}\right)^{3/2} - \frac{3 D}{16 \pi^2 \phi^4}
\right] \label{Spot1}.
\end{equation}
where $C_{3/2}$ is a positive constant and $D = 2 a_0^4 T_3$. Remarkably, this potential is same as that obtained in \cite{Bau2,Bau3,KP} after the volume modulus is stabilized at the instantaneous minimum by making use of the adiabatic approximation. However the interpretation and the constraint on $C_{3/2}$ are different. The fact that the generic form of the potential is same, it tells us that the holographic point of view is another approach to compute the inflation potential.  In the
next section, we discuss the constraints on allowed values of the
model parameters which can give rise to an inflationary dynamics
consistent with observation.
\section{Inflationary potential and fine tuned parameters}

 The possibility of a viable D-brane inflation based upon
 the effective potential (\ref{Spot1}) was examined in Ref.\cite{acps}. In what follows, we shall briefly review
 the analysis for the dynamics of inflation presented there. For convenience,
we use the following dimensionless form of the
 potential,
\begin{eqnarray}
\label{Spot2}
 &&{\cal V}={\cal
D}\left(1+\frac{\alpha^2}{3}x^2-C_{3/2}x^{3/2}-\frac{3 {\cal D}}{16
\pi^2  x^4}\right), \\
&& x=\phi/M_{UV},  {\cal V}=V/M_{UV}^4, \nonumber\\
&& {\cal D}=D/M_{UV}^4, \alpha=M_{UV}/M_P .\nonumber
\end{eqnarray}
 In  this scenario, $0< x < 1$ as $\phi<
\phi_{UV} = M_{UV}$ and the mobile D3-brane moves towards the ${\bar
D}$3 brane located at the tip of the throat corresponding to $x=0$.
The slow roll parameters for the generic field range are given by
\begin{eqnarray}
&&\epsilon= \frac{1}{2 \alpha^2}\left(\frac{{\cal V}_{,x}}{{\cal
V}(x)}\right)^2 \simeq \frac{1}{2 \alpha^2}\Big[
\frac{2\alpha^2}{3}x-\frac{3 C_{3/2}}{2} x^{1/2} \nonumber \\
&&+\frac{3 {\cal D}}{4 \pi^2 \alpha^4 x^5}\Big ]^2 \\
&&\eta=\frac{1}{\alpha^2}\frac{{\cal V}_{,xx}}{{\cal V}(x)}\simeq
\frac{2}{3}-\frac{3C_{3/2}}{4\alpha^2}\frac{1}{x^{1/2}}-\frac{15{\cal
D}}{4 \pi^2 \alpha^6 x^6}\label{eta1}
\end{eqnarray}
 Since in the present case $|\epsilon| < |\eta|$ , it is sufficient
to consider $\eta$ for discussing the slow roll conditions. It
follows from Eq.(\ref{eta1}) that $\eta$ is always less than one; it
decreases as $x$ moves towards the origin. At a particular value of
$x$, the slow roll parameter $\eta = - 1$ marks the end of inflation
and thereafter it takes large negative values  as $x \to 0$.

Note that in the present set up, the field $x$ rolls from $x=1$
towards $x = 0$ where $\Bar{D}3$ brane is located. This implies that
the field potential should be monotonously increasing function of
$x$.
However, depending upon the numerical values of the model parameters
(${\cal D},~C_{3/2}~\alpha$), the effective potential may acquire a
minimum with large value of
 $\eta_{min}$ giving rise to spectral index inconsistent with
observation. It is observed that the region where $\eta$ is small
lies below $\phi_{min}$, in this case, and is not accessible
dynamically.
 The choice of
parameters should also be consistent with the constraints coming
from the compactification as well as other considerations necessary
for the validity of the effective potential
 (\ref{Spot1}).
Monotonicity of ${\cal V}(x)$ is ensured
provided that ${\cal V}_{,x}>0$,
\begin{equation}
C_{3/2}<\frac{2}{3}\left(\frac{2\alpha^2\sqrt{x}}{3}+\frac{3 {\cal
D}}{4 \pi^2x^{11/2}}\right).
 \label{C3by2}
\end{equation}
In Eq.({\ref{C3by2}), the right hand side as a function of $x$ has
minimum at $x=x_{min}=(99 {\cal D}/8 \pi^2\alpha^2)^{1/6}$  which
imposes a constraint on the coefficient, $C_{3/2}$,
\begin{equation}
C_{3/2}<\frac{8\times
2^{3/4}}{3^{5/6}11^{11/12}}\left(\frac{\alpha^{22} {\cal
D}}{\pi^2}\right)^{1/12}.
\end{equation}

\begin{figure}
\begin{center}
\includegraphics[angle=-90,width=9cm]{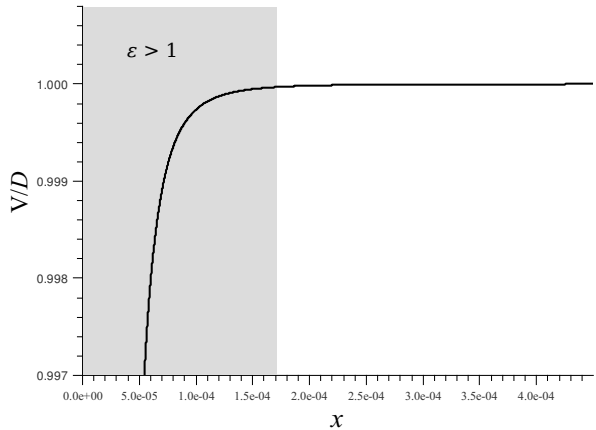}
\caption{Plot of the effective potential  for
$\alpha=0.5$, $C_{3/2}=4.50272\times 10^{-3}$, ${\cal
D}=1.3820\times 10^{-18}$. The potential is extremely flat in the
field range of interest  and becomes large negative as $x\to 0$. The
shaded region shows the non-inflationary part of the potential.}
\label{pot}
\end{center}
\end{figure}
The potential (\ref{Spot2}) is plotted in (\ref{pot}) for $0<x<1$
keeping in mind the aforementioned constraints. The parameters
should therefore be chosen such that the potential has the right
behavior. It turns out that for ${\cal D}^{1/4}\lesssim 10^{-4}$ and
$\alpha <1$, the observational constraints can be satisfied by
pushing $C_{3/2}$ towards numerical values much smaller than one,
(see Table \ref{table}). Since the field range viable for inflation
is narrow, the potential should be made sufficiently flat to obtain
required number of e-foldings. In particular, it means that the slow
roll parameter $\epsilon$ is very small leading to low value of
tensor to scalar ratio of perturbations. This feature, however,
becomes problematic for scalar perturbations. Indeed, since
$\delta_H^2 \propto {\cal V}/\epsilon$ and ${\cal V} \sim {\cal D}$,
smaller values of $\epsilon$ lead to larger values of density
perturbations. We also emphasize a peculiarity of the potential
associated with the last term in (\ref{Spot2}). The constant ${\cal
D}$ not only appears as an overall scale in the expression of the
effective potential, it also effects the behavior of $V(x)$ in a
crucial manner, for instance it changes the slow roll parameters,
which makes it tedious to set the COBE normalization.

Our numerical investigations, considering the allowed values of
${\cal D}$ , $\alpha$ and $C_{3/2}$ coming from the validity of the
model considered here,  confirm that there exist regions in
parameter space which allows us  to satisfy all the observational
constraints including the COBE normalization. For instance, one of
the choices is given in Table \ref{table}. We note that ${\cal D}$,
which is the scaled value of $D$ as defined in (\ref{Spot2}) is
related to the warped tension of the D3-brane. The warp factor at
the tip of the throat, in turn, is an exponential function of the
integer used in the compactification scheme . It is, thus, permitted
to tune them to obtain the quoted values of $\cal D$ in Table
\ref{table}. Similarly, $C_{3/2}$ which is a parameter corresponding
to the strength  of the perturbation in dual CFT is also very small.

 However, it is also possible to allow small changes in the quoted values of these
parameters and still satisfy the observational constraints. We
should remark here that  the model parameters should be suitably
fine tuned to satisfy the observational constraints. For instance,
$C_{3/2}$ is required to be fine tuned to the level of one part in
$10^{-7}$. The changes at the seventh decimal puts physical
quantities out side their observational bounds otherwise the
potential can develop a local minimum and hence can spoil all the
nice features of the model. Similarly other parameters are also
needed  to be fine tuned. For example, if we take ${\cal
D}=1.210\times 10^{-17}$ instead of ${\cal D}=1.211\times 10^{-17}$,
the field gets into the fast roll regime before we could meet the
requirements of having $60$ e-folds as well as being consistent with
COBE normalization. The parameter $\alpha^{-1}$ also requires fine
tuning of the order of one part in $10^{-5}$. It is quite possible
that the systematic search of these parameters based on Monte-Carlo
method discussed in Ref.\cite{HC} might help to alleviate the fine
tuning problem. It is, nevertheless, remarkable that the correction
to the inflaton potential without changing the throat geometry, as
being discussed here but originally reported in Ref.\cite{BDKKM},
allows not only to solve the well know $\eta$ problem  but also
helps in satisfying all the observational constraints given by the
WMAP5.}

\begin{figure}
\begin{center}
 \includegraphics[angle=-90,width=9.5cm]{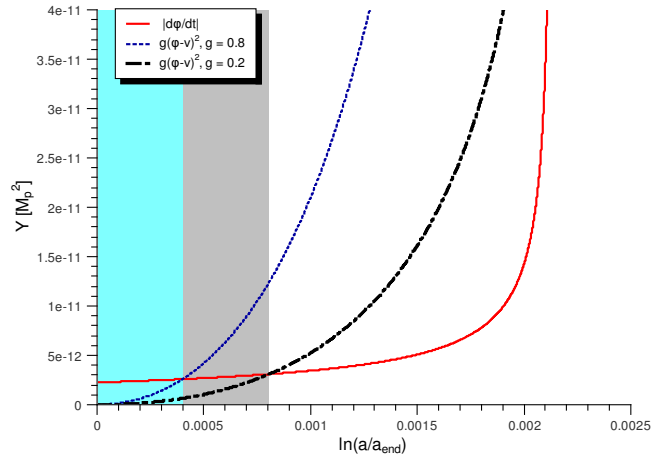}
 \caption{Post inflationary evolution of $g(\phi-v)^2$ and
 $|\dot{\phi}|$ versus $\ln(a/a_{end})$. Reheating takes place
 around $\phi=v$.
 The violation of adiabatic condition takes place in the
  shaded regions. }
  \label{adiabat}
\end{center}
\end{figure}
Let us note that in a viable inflationary scenario, inflation should
be followed by a successful reheating. In the model under
investigation, the D-brane effective potential, with parameters appropriately chosen,  does not posses
a minimum and therefore, the conventional reheating mechanism  does
not work in this case. The instant preheating mechanism might be
helpful in such a case. At the end of inflation, the effective
potential becomes steep and assumes large negative values as the
field approaches the origin. The perurbative string theory
calculations used to obtain the effective potential (\ref{Spot2})
become unreliable near the tip of the throat. The non-perturbative
result might give rise to an effective potential with a minimum or
the brane-brane collision might lead to particle production.
However, if the true D-brane potential is run away type, the instant
preheating mechanism might give rise to a viable thermal history of
universe. In the next section, we shall investigate the said
possibility.

\bigskip
\begin{table}
 \begin{tabular}{|c|c|c|c|c|}
 \hline
$\alpha$ &  $C_{3/2}$ &   $\mathcal{D}$        &  $n_s$  &$\delta_H^2$\\
\hline
$0.50$ & $4.50272\times 10^{-3}$ & $1.3820 \times 10^{-18}$    & $0.96$  &$2.4\times 10^{-9}$ \\
$0.25 $& $1.58407\times 10^{-3}$ & $2.0829 \times 10^{-17}$ &0.96 &$2.4\times10^{-9}$\\
$0.10 $  &$4.00285\times 10^{-4}$&  $8.0254\times 10^{-16}$  &0.96
&$2.4\times10^{-9}$\\
 \hline
\end{tabular}
\caption{ A possible  choice of parameters which can give rise to a
viable inflation.} \label{table}
\end{table}
\bigskip
\section{D-brane inflation followed by instant preheating}
Before getting into instant preheating, let us estimate the
reheating temperature due to gravitational particle production. The
space time geometry undergoes a crucial transition at the end of
inflation involving essentially a non-adiabatic process which leads
to particle production. The energy density of radiation produced in
this process is given by\cite{ford,sp,lidsey},
\begin{equation}
\rho_r\simeq 0.01 \times g_p H_{end}^4,
\end{equation}
where $g_p$ is the number of different species produced at the end
of inflation and varies typically between $10$ and $100$. For a
possible choice of inflationary parameters given by: ($\alpha=0.1$,
$C_{3/2}=4.00285\times 10^{-4}$, ${\cal D}=8.0254\times 10^{-16}$),
we find that
\begin{equation}
H_{end}=1.636\times 10^{-10} M_p \Rightarrow ~~T \simeq \rho_r^{1/4}
\simeq 4 \times 10^8 GeV.
\end{equation}
The low reheating temperature signifies that the gravitational
particle production is an inefficient process. We have verified that
temperature does not improve significantly for other possible
choices of inflationary parameters. The instant preheating generally
gives rise to much higher reheating temperature than the
gravitational particle production. In this scenario, in order to
achieve reheating after inflation, one assumes that the inflaton
$\phi$ interacts with another scalar field $\chi$ which has a
Yukawa-type interaction with a Fermi field $\psi$. The interaction
Lagrangian has the following form\cite{FKL,FKL1,sahnisami,R,chiba}
\begin{equation}
 L_{int} = -\frac{1}{2}g^2(\phi-v)^2\chi^2 - h\bar{\psi}\psi\chi,
 \label{Lag}
\end{equation}
where $v$ is chosen corresponding to the vanishing of $m_\chi$ at
the end of inflation $(v=\phi_{end})$. This is done keeping in mind
that the field $\chi$ has to be extremely light so that it does not
participate in the dynamics during the  inflationary epoch,  which
is one of the requirements of the proposal of instant reheating. In
the present context, where all the moduli are stabilized and we are
considering the scenario of large volume compactification, the
scalar field $\chi$ appearing in the Lagrangian (\ref{Lag}) could be
thought of as a low lying Kaluza-Klein excitation so that it is
automatically light. \footnote{We thank Ashoke Sen for a discussion
clarifying this point.}.

\begin{figure}
 \includegraphics[angle=-90,width=9.5cm]{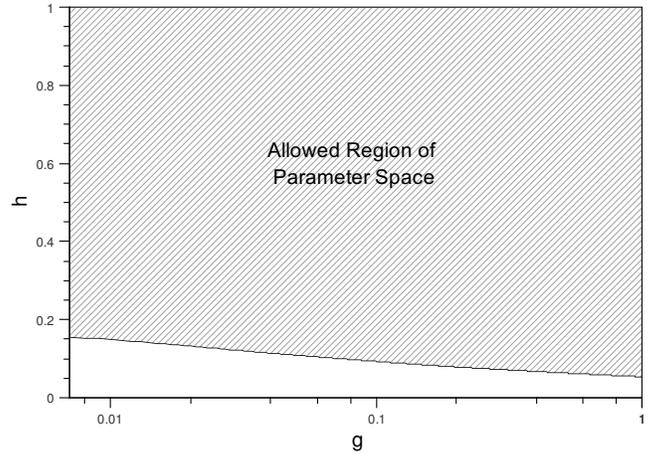}
\caption{Plot shows the region in parameter space $(g,h)$
corresponding to $\alpha=0.5$ allowed by post inflationary
non-adiabatic process and the consideration of back reaction. The
upper limit on $h\&g$ is dictated by the validity of perturbation
theory. The shaded region $0.006 \lesssim g \lesssim 1$ and $0.053
\lesssim h \lesssim 1$
 leads to successful instant reheating  }
\end{figure}
The preheating commences when $m_{\chi}=g|\phi-v|$ starts changing
non-adiabatically after inflation had ended,
\begin{equation}
|\dot{m_{\chi}}| \gtrsim m_{\chi}^2 \qquad \textrm{or} \qquad
|\dot{\phi}| \gtrsim g (\phi-v)^2.
\end{equation}
In our model, the particle production takes place when $\epsilon
\approx 1 $ and $\dot{\phi}\ne 0$. The effective mass of
$\chi$ grows starting from zero ($|\phi-v| = 0$)during the post
inflationary evolution. The adiabatic condition is satisfied at the
end of inflation when $m_\chi=0$ but holds no longer
thereafter. For generic values of the parameter $g$, we confirm from
the numerical calculations that the term $g(\phi-v)^2$ grows
faster than $|\dot{\phi}|$ in the beginning ($|\dot{\phi}|$
will grow faster near singularity). The condition for
particle production is satisfied provided that,
\begin{equation}
\phi\lesssim \phi_{pro}=v-\sqrt{\frac{|\dot{\phi}_{pro}|}{g}}.
\end{equation}

The region of the violation of the adiabaticity is bounded by the
intersection point of two curves, $|\dot{\phi}|$ $\&$ $g(\phi-v)^2$
. Larger $g$ means smaller time duration of particle production and
the corresponding  larger value of $|\phi_{pro}|$.
In Fig.\ref{adiabat}, we have displayed the post inflationary
evolution of $|\dot{\phi}|$ and $g(\phi-v)^2$. We find that for any
generic value of the parameter $g$, the preheating takes place
instantaneously. The lower limit of $g$
 is fixed such that the non-adiabatic process stops before
 singularity is reached.
For instance, the
 variation of $g$ is given by,
 $0.006\lesssim g\lesssim 1$ in case of $\alpha=0.5$.
\begin{figure}
 \includegraphics[angle=-90,width=9cm]{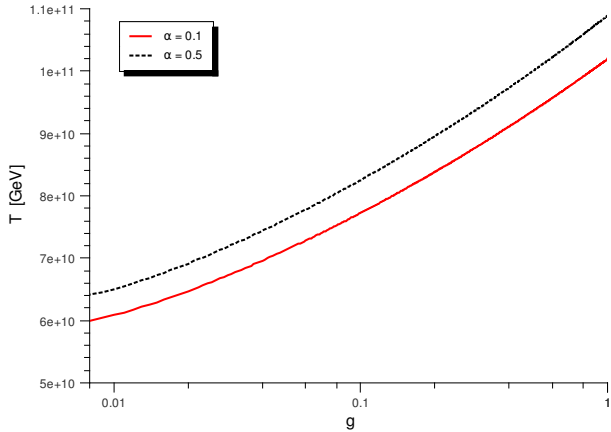}
 \caption{Plot of reheating temperature versus $g$ for two different
 values of $\alpha$. The plot shows that the reheating temperature, $T \simeq \rho_\chi^{1/4}$
 varies between $6\times10^{10}-1.1\times10^{11}$ GeV depending upon the
 allowed values of the parameter $g$.}
\end{figure}
We can estimate the momentum of the produced particles  using the
uncertainty relation
\begin{equation}\label{eq:kp}
k_p \sim {\delta t}^{-1} \sim \frac{|\dot{\phi}|}{|\phi|}
=\alpha \left | \frac{y}{x} \right| M_{p},
\end{equation}
where $y=\dot{\phi}/M_{UV}^2$.
The occupation number of $\chi$ particles assumes a finite value
during the time interval $\delta t$,
\begin{equation}
 N_k \approx \exp (-\pi k^2/k_p^2),
\end{equation}
which allows us to  estimate the $\chi$ particle number density,
\begin{equation}
 n_{\chi} = \frac{1}{2\pi^2} \int^{\infty}_{0} k^2 N_k dk \approx \frac{k_p^3}{8\pi^3}.
 \label{nchi}
\end{equation}
Using Eq.(\ref{nchi}), we can finally compute the energy density of
the $\chi$ particles,
\begin{equation}\label{eq:rhochi}
 \rho_{\chi} = m_{\chi} n_{\chi} \left ( \frac{a_{end}}{a} \right )^3
=g|(\phi-v)| \left ( \frac{k_p^3}{8\pi^3} \right ) \left (
\frac{a_{end}}{a} \right )^3,
\end{equation}
where $\phi$ is $\phi_{prod}$ which is obtained from the
intersection of two curves $g(\phi-v)^2$ and $|\dot{\phi}|$. It
would be convenient to express the $\rho_{\chi}$ through the
dimensionless variables $x,y$ making use of $(\ref{eq:kp})$ $\&$
$(\ref{eq:rhochi})$ as
\begin{equation}
 \rho_{\chi} =\frac{g}{8\pi^3}|x-{\tilde v}|\alpha^4\left | \frac{y}{x} \right |^3
M_p^4 \left ( \frac{a_{end}}{a} \right )^3,
\end{equation}
where ${\tilde v} = v/M_{UV}$.
We assume that the energy of $\chi$ particles which are produced
 after inflation, is  thermalized  instantaneously giving rise to the radiation energy density,
\begin{equation}
 \rho_{r} = n_{\chi} m_{\chi}
= \frac{g\alpha^4}{8\pi^3}\left | \frac{y^3(x-{\tilde v})}{x^3} \right |
M_p^4,
\end{equation}
where we have dropped the cosmic dilution factor due to instant
character of reheating process.
Our numerical calculations show that the reheating temperature,
$10^{10} GeV\lesssim T \lesssim 10^{11} GeV$ (for permissible values
of the parameter $g$) which is higher by about three orders of
magnitudes compared to its counter part associated with
gravitational particle production.

The decay rate of $\chi$ particles  to $\psi$ and $ \bar{\psi}$ is
given by, $\Gamma_{\bar{\psi}\psi} = h^2m_{\chi}/8\pi$ which must be
larger than the expansion rate of universe for the back reaction of
$\chi$ particles, on the background post inflationary evolution of
$\phi$, to be negligible .
This gives the lower bound on the parameter $h$,
\begin{equation}
\Gamma_{\bar{\psi}\psi} > H \Rightarrow h^2 \gtrsim 8\pi\frac{
H_{end}}{g |\phi_{prod}-v|}. \label{hrange}
\end{equation}
where we have made use of the fact that the Hubble parameter,
$H\simeq H_{end}$ in the  non-adiabatic region. Relation
(\ref{hrange}) along with generic values of parameter $g$ leads to
the region in parameter space $(h,g)$ which may give rise to viable
instant preheating.

\section{Conclusions}
In this paper we have investigated a possibility for reheating
followed by D-brane inflation described by the effective potential
(\ref{Spot2}). The D-brane effective potential belongs to the
category of {\it non-oscillatory} models for which the conventional
reheating mechanism does not work. The gravitational particle
production is generally inefficient and leads to reheating
temperature around $10^8 GeV$  in the present model. The instant
preheating is demonstrated to be more efficient than gravitational
particle production. The mechanism involves interaction of inflaton
with a scalar degree of freedom $\chi$  with zero bare mass $ \left
( m_\chi^2=g^2(\phi-v)^2 \chi^2 \right )$. The effective mass of
$\chi$ linearly depends on the inflaton field which is identically
zero at the end of inflation but increases thereafter. The scalar
degree of freedom interacts with fermion and anti-fermion pair
($h\chi \bar{\psi} \psi$). We found a viable range of parameters for
preheating demanding the fulfillment of non-adiabaticity condition
during post inflationary evolution and at the same time keeping the
back reaction under control. This requirements leads to the range of
parameters specified by: $0.006\lesssim g\lesssim 1$ and $
0.053\lesssim h \lesssim 1$. The corresponding preheating
temperature $T$ is shown to be in the range : $10^{10} GeV\lesssim T
\lesssim 10^{11} GeV$. We find that, for the permissible values of
parameters ($h\& g$), the process of preheating is almost
instantaneous. Interestingly, the allowed range of parameters $h$
and $g$ and the corresponding preheating temperature turn out to be
weakly dependent on the choice of numerical values of inflationary
parameters suitable to observations.

To be rigorous, we should note that the number of e-folds ${\cal
N}$, in general, depends upon the reheating temperature. In our
analysis, we have assumed horizon exit corresponding to 60 e-folds
which is a popular choice. However, the general constraints do not
allow  ${\cal N}$ to take appreciably larger values than the one
quoted here. We have observed that generic variations of ${\cal N}$
does not lead significant improvement of the preheating temperature.

 We, therefore, conclude that instant reheating is a
viable mechanism for a single throat model.

Let us  point out here an important issue related to the energy
transfer from inflaton to the standard model particles. Though
concrete proposal for such mechanism does not exist at present, one
possibility could be to embed the flux compactification of Type IIB
theory in F-theory compactification with fluxes.  One could then
consider the scenario of intersecting branes  to obtain the spectrum
of particles with standard model gauge group. The fermion field that
we have considered in this paper could very well be one of the
fields in the standard model living on the world volume of the
intersecting branes. Another possibility is related to the
assumption that the standard model particles are localized in one of
the D7-branes wrapping the four cycle of the compact internal
manifold used to generate the non-perturbative potential. Such a
possibility is recently analyzed in \cite{ko} (see also
Ref.\cite{others} on the related theme). It should be noted that in
a realistic scenario, the energy transfer to standard model degrees
of freedom may not be that efficient as assumed in the present
analysis while comparing the instant preheating with gravitational
particle production. In our opinion, this is an important issue
which deserves further investigations.
\section{Acknowledgements}
 We thank Zini Rahman for useful comments. IT is supported by ICCR fellowship.
 MS is supported by Indo-Japan project (Grant No:
 DST/INT/JAP/P-83/2009). SP thanks Centre for theoretical physics,
 Jamia Millia Islamia, New Delhi for hospitality.

\end{document}